# From Theoretical Foundation to Invaluable Research Tool: Modern Hybrid Simulations

D. Krauss-Varban

*Space Sciences Laboratory, UC Berkeley, Berkeley, CA, USA*

In many plasmas, in particular in space science, protons govern much of the essential physics. Minority ions, suprathermal tails, and electrons at times account for additional details. However, electron effects usually appear on much smaller spatial and temporal scales. For more than two decades, scientists have refined computational models that concentrate on the dominating and larger-scale ion kinetic physics, while treating the much lighter electrons as a charge-neutralizing fluid. These physics-based, algorithmic model descriptions are called *hybrid codes*, which have established an invaluable position for themselves – clearly distinct from MHD calculations, but simpler and much more efficient than full-particle simulations. Here, the foundation and numerical details of hybrid codes are reviewed, and differences between existing algorithms are elucidated.

## 1. Introduction

Hybrid codes have been around for more than 30 years [Harned, 1982; Winske, 1985]. They derive their importance from the fact that in many plasma systems, in particular in space science, ions and their kinetic effects dominate much of the physics. While electrons often account for additional (and at times, even essential) details, such details generally appear to ride on top of the proton physics, on much finer spatial and temporal scales. Given that kinetic (particle) simulations are much more expensive than fluid simulations, it comes as no surprise that for more than two decades, scientists have elaborated on models that concentrate on the larger-scale ion kinetic physics, while treating the much lighter, presumably less important and shorter-scale electrons simply as a charge-neutralizing fluid. Much experience has been gained over time on the numerical properties of hybrid codes. In recent years, the vast majority of published computations have used one of three distinct types: a direct solver (one-pass method) [Thomas et al., 1990; Fujimoto, 1990; Omidi et al., 2001], the predictor-corrector scheme [Harned, 1982; Quest, 1989], and algorithms based on the moment method [Quest, 1989; Matthews, 1994]. Here, we give an overview of the foundation of hybrid codes and describe in detail the differences of existing approaches with demonstrated relevance and success. We conclude with a brief performance comparison. The history and numerics of hybrid codes are further detailed in Omidi et al. [2001] and in Winske et al. [2003].

## 2. Foundation

Hybrid codes are similar to other particle-in-cell (PIC) codes. As such, the reader may be referred to one of the many good books and review articles on PIC algorithms.

The codes used in any significant number of application to date employ, like most particle simulations, a leapfrog method to advance the spatial and velocity components of the pseudo-particles. This means that the spatial positions and velocities are calculated on a temporal grid ½ time step apart. From this, moments of the ion (charge) density and currents are obtained to allow calculation of the electric and magnetic field.

The magnetic field is simply advanced from Faraday's law

$$\delta \mathbf{B} / \delta t = -c \, \nabla \times \mathbf{E}, \quad (1)$$

involving the electric field as a source at a certain, defined time step. The main difference to full-particle codes is that the electrons are described as a charge-neutralizing fluid. Consequently, no electron pseudo-particles are advanced, and no electron moments are evaluated. Instead, an approximate form of the electron momentum equation is utilized. The form most often used only retains the $\mathbf{j} \times \mathbf{B}$ force, the gradient of a scalar pressure $p$, and an explicit resistivity term $\eta \, \mathbf{j}$. When the current and ion velocity moment are substituted for the electron velocity, the electron momentum equation takes the form of a generalized Ohms law based on known quantities:

$$\mathbf{E} = -\mathbf{v}_i \times \mathbf{B}/c + \mathbf{j} \times \mathbf{B}/(qnc) - \nabla p_e/(qn) + \eta \, \mathbf{j} \quad (2)$$

Other variations of the hybrid code employ a tensor description of the electron pressure and retain a finite electron mass term [Winske et al., 2003]. Here, however, we only describe conventional hybrid codes, which concentrate on the ion scales. Note that eq. 2 is a state equation. This means that results for $\mathbf{E}$ can be obtained without time advance, from a given set of sources (density $n$, ion velocity/current $\mathbf{v}$, and $\mathbf{B}$) at a certain time. The most important point here is to realize that due to the use of a leapfrog method for the particles, the two desired source quantities ($n$, $\mathbf{v}$) will not be available at the same time step when solving for $\mathbf{E}$. Further, advancing $\mathbf{B}$ in a scheme that is more accurate and more stable than simple forward differencing requires knowledge of an approximate $\mathbf{E}$ at a later time, that is yet to be calculated (and that implicitly involves $\mathbf{B}$ and the moments at this later time). Conversely, calculation of the moments at such advanced times may require pushing the particles using $\mathbf{E}$ and $\mathbf{B}$ at advanced times. The existing hybrid code variants differ in their approach on how to remedy this interdependence in a stable, accurate, and fast way.

## 3. Hybrid code variants

The basic codes mentioned above are described in detail in the literature. For here, it suffices to summarize their general guiding principles; see Fig. 1 for a sketch of a simple explicit solver, and Fig. 2 for a predictor-corrector scheme. The direct solve (one-pass method) tries to avoid a second particle push to align the sources for **E**. Instead, it uses an extrapolation of the velocities when calculating **E**. For this reason, it is a relatively fast algorithm.

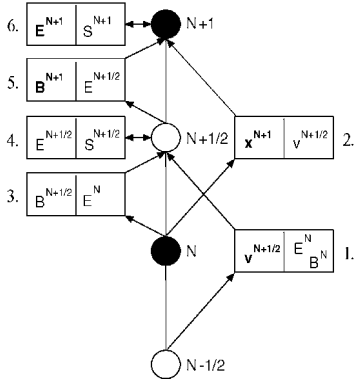

Fig. 1. Explicit hybrid algorithm. Bold letters: permanent solution, regular letters: temporary values. The left of each box shows calculated quantity, the right its sources. S stands for **B**, $n$, and $v_i$, combined.

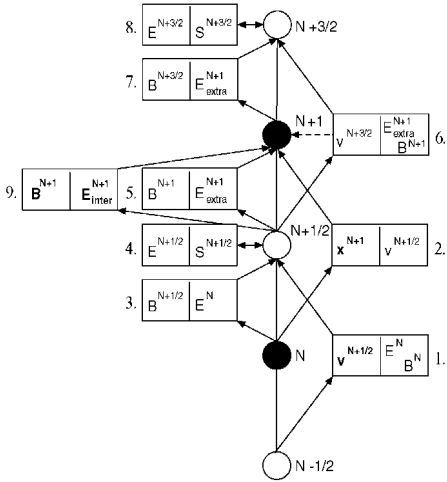

Fig. 2. Predictor-corrector algorithm. Because of symmetry between forward and backward differences, steps 5, 7, 9 become time-centered.

The predictor-corrector scheme is known for its accuracy in describing the shorter wavelength/higher frequency portion of the wave spectrum. It consists of a two-step process, in which an extrapolated value of **E** is used to advance a temporary **B** and to move the particles for a second moment collect – while a corrector step ensures accurate sources for the final **E**-calculation, maintaining a time-centered philosophy.

The CAM-CL method, as developed by Mathews [1994], avoids the second particle push by employing a moment method similar to earlier variations of the hybrid code [Quest, 1989]. The equations involved are somewhat cumbersome, but in the end there is little overhead compared to the one-pass method.

When extrapolating the velocity or density moments to match desired points in time, one can use higher-order schemes [Fujimoto, 1990; Thomas et al., 1990]. However, in extensive tests and comparisons, we have found no advantage of doing so. Thus, we only report here on using linear extrapolation/ interpolation in the moments for both the proper time-slice evaluation, and for evaluating moments within sub-stepping.

For very high frequency/ short wavelength waves, one can make the argument that typically, only modes that are almost field-aligned are not heavily damped and contribute to the physical and numerical properties of the simulation. These modes do not involve significant perturbations of the density and velocity components. As such, it is advantageous to sub-step the electric field solver and magnetic field advance without recalculating the particle sources [Swift, 1995; Mathews, 1994]. We use a Runge-Kutta [Press et al., 1992] scheme for this sub-stepping in both the one-pass and CAM-CL codes. Moreover, the Runge-Kutta scheme is generally used in our one-pass algorithm instead of the two-step explicit process indicated in Fig. 1. In the predictor-corrector code, it is more practical to continue the predictor-corrector scheme on a sub-step basis – which is what we do. The particle push/ moment collect indicated in steps 1,2, and 6 of Fig. 2 are then only executed in the first and last sub-step, respectively, and the "N+3/2" point refers to sub-stepped times, getting increasingly closer to N+1 when using a large number of sub-steps. In this code, moments are interpolated/ extrapolated as needed for specific sub-step points. As a guideline, we find that for typical plasma parameters and numbers of particles per cell, sub-stepping somewhere beyond 4 to 16 per particle time step is less efficient than using a smaller time step for the entire computation. The exact number depends on the cost of the particle push compared to the field solve (i.e., on the number of particles per cell, and on field solver overhead incurred during parallel processing).

Coming back to the question of at what particle position the velocity moments should be evaluated, there are two competing principles. Standard practice (1) is to honor the fact that the velocities are evaluated ½ temporal step before the positions, and thus to use particle positions for the **v**-moment evaluation at this time. A simpler method (2) uses the particle positions that are more readily available, namely, ½ half step farther advanced, after the particle positions have been updated, and thus, at the same time the density moment is collected. It turns out that method (2) is susceptible to an instability transverse to **B** reminiscent of mirror waves under certain parameter combinations (e.g., when the density is low and the magnetic field is large, as in the magnetosphere). On the other hand, method (1) can lead to artificial particle heating when the streaming velocity $|v_0|$ of the plasma is large compared to the thermal velocity $v_{th}$, unless a very small time step is chosen. Then, in many codes, instead of the usual Courant-Friedrichs-Levy (CFL) condition on the particles, $\Delta t < \Delta x / v$, a time step $\Delta t < (v_{th} / |v_0|) \Delta x / v$ is required. Here, $v$ stands for a typical combination of thermal and streaming velocities, and the conventional CFL condition ensures that not too many particles cross a cell in a time step. Note that here is a similar condition on the fastest waves in the system (see below).

To resolve the numerical challenges outlined above, in the one-pass code we employ a method based on the local and instantaneous CFL condition of the particle (the ratio of its spatial advance in one time step, compared to the cell size). In other words, a particle that is fast in a particular direction is evaluated close to method (2), whereas a slow-moving particle approximately uses method (1). It is more difficult to employ such a scheme in the CAM-CL algorithm, which in its moment prediction uses method (1) by default. In our tests, the predictor-corrector algorithm does not benefit from this modification, and works best with method (1).

## 4. Comparison

In extensive 2-D tests, we find that for most practical circumstances, any of the above three codes give excellent results (see, e.g., the comparison in Karimabadi et al. [2004]). The CAM-CL code appears to be stable at slightly longer time steps, even when the CFL condition on the fastest waves is marginally violated. In our tests, it also gave temperatures that are slightly too low. These two characteristics indicate that this code is more diffusive than the others – which is expected, given its underlying philosophy. We have not found any significant differences concerning the out-of-plane magnetic field or other important plasma parameters, when run at a sufficiently small time step. Differences between the codes surface in challenging situations, where excellent convergence at the least amount of CPU time is sought.

At a sufficiently small time step, all three algorithms have negligible numerical diffusion/resistivity. Even very thin current sheets with thickness small compared to the ion inertial length will not reconnect unless resistivity is (locally) added, breaking the frozen-in condition for the electrons. However, in anti-parallel magnetic field configurations embedded in streaming plasmas, numerical diffusion shows its presence and reconnection may occur. In this case, we found that the CAM-CL code has by far the largest amount of diffusion, followed by the one-pass method, with the predictor-corrector being by far the least diffusive.

The predictor-corrector algorithm uses more CPU time due to its second particle advance. However, when implemented as indicated in Fig. 2 (only the **v**-moments are evaluated a second time, not $n$), it is only 20% slower than the one-pass method, under typical circumstances. Having **v** available at N+1 rather than only at N+3/2 simplifies calculating interpolated values in the sub-stepping process. In practice, to preserve the non-linear properties of the particle push, this is done by advancing to N+3/2 and then dividing the advance by two. Extrapolation to N+3/2 then simply recovers the full step.

## 5. Summary

Hybrid simulations have been used in 1-D through 3-D domains for a vast array of topics in space plasmas and other areas and applications. They have been successfully employed extending from very local simulations of instabilities, current layers, and discontinuities to global simulations of planetary magnetospheres. Several algorithm variants exist and have demonstrated their capabilities. Of these, the one-pass method is simple, accurate, and efficient, and is often used as a standard code. The CAM-CL method is equally accurate in most applications, but is more diffusive and may (at least in its standard formulation) cause significant numerical magnetic field line reconnection in a moving plasma. Under circumstances where best convergence and conservation properties are required, the predictor-corrector method outshines the other schemes. Modern versions of the predictor-corrector algorithm have little overhead compared to direct methods, and compete well when high accuracy is needed and/or long run-times are required.

**Acknowledgments** This work was in part supported by NSF CISM, which is funded by the STC Program of the NSF as agreement ATM-0120950. Other support derived from NSF grant ATM-0454664 and NASA grant NNG04GH38G.